\documentclass[12pt]{article}
\usepackage{hhline}
\usepackage{latexsym,graphicx}
\usepackage{subfigure}
\usepackage{amssymb}
\usepackage{amsmath}
\usepackage{amscd}
\usepackage{amsthm}
\usepackage{float}
\usepackage[left=2cm,top=2.5cm,right=2.5cm,bottom=1.5cm]{geometry}   
 
\begin{document}
\begin{center}
\large{\bf{Crossing the phantom divide line in universal extra dimensions}} \\
\vspace{10mm}
\normalsize{Nasr Ahmed$^1$,$^2$ and Anirudh Pradhan$^3$}\\
\vspace{5mm}
\small{\footnotesize $^1$ Mathematics Department, Faculty of Science, Taibah University, Saudi Arabia.} \\
\small{\footnotesize $^2$ Astronomy Department, National Research Institute of Astronomy and Geophysics, Helwan, Cairo, Egypt\footnote{nasr.ahmed@nriag.sci.eg}}\\
\small{\footnotesize $^3$ Department of Mathematics, Institute of Applied Sciences \& Humanities,
GLA University, Mathura-281 406, Uttar Pradesh, India}
\end{center}  
\date{}
\begin{abstract}
We investigate the cosmic acceleration and the evolution of dark energy across the cosmological constant boundary in universal extra dimensions UED. We adopt an empirical approach to solve the higher-dimensional cosmological equations so that the deceleration parameter $q$ is consistent with observations. The expressions for the jerk and deceleration parameters are independent of the number of dimensions $n$. The behavior of pressure in $4$D shows a positive-to-negative transition corresponding to the deceleration-to-acceleration cosmic transition. This pressure behavior helps in providing an explanation to the cosmic deceleration-acceleration transition although the reason behind the transition itself remains unknown. In the conventional $4$D cosmology, there is a no-go theorem prevents the EoS parameter of a single perfect fluid in FRW geometry to cross the $\omega=-1$ boundary. The current model includes a single homogenous but anisotropic perfect fluid in a homogenous FRW metric with two different scale factors in the ordinary $4$D and the UED. In contrast to the conventional $4$D cosmology, we have found that the dark energy evolution in UED shows $\omega=-1$ crossing. however, the no-go theorem is still respected in $4$D where the EoS parameter doesn't cross the $\omega=-1$ boundary.

\end{abstract}
\section{Introduction and motivation}

The discovery of the accelerating expansion of the present universe is a remarkable development in modern cosmology \cite{11,13,14}. One way to explain this late-time cosmic acceleration is by modifying (or even replacing) general relativity \cite{39, noj8, torsion, 1}. Another way is by assuming the existence of energy component with negative pressure (dubbed as dark energy DE) which gives a repulsive gravity effect \cite{quint, chap, phant, ess, tak, ark}. On the other hand, the idea of the existence of some hidden extra dimensions in our universe has played a fundamental role in modern theoretical physics. examples of such extra-dimensional theories include: Kaluza-Klein theory \cite{kk} where gravity and electromagnetism can be unified by adding one extra dimension, the $11$D supergravity \cite{ss}, the $10$D superstrings \cite{sss}, and the $11$D M-theory which is simply supergravity with two boundaries \cite{nasrwork}. The $11$D M-theory can be reduced to $5$D system with two boundaries. As a result of this dimensional reduction, new elementary particles arise which makes the resulting $4$D effective theory very interesting phenomenologically. In general, dimensional reduction leads to an infinite tower of Kaluza-Klien states perceived in 4D as new massive particles. This is relevant to cosmology as it gives a possibility for a natural candidate of dark matter. If the lightest Kaluza-Klien particle (LKP) is stable, it would still exist in the present universe and can possibly possess all the properties of the WIMP (weakly interacting massive particles) dark matter if it is electrically neutral and non-baryonic \cite{naskk}. The reduction from $5$D to our observed $4$D world can be obtained by integrating out the 5th extra dimension \cite{nasrwork}. Some recent interesting proposals of extra-dimensional theories include: brane-world scienarios \cite{brane} which describe the universe as a $3 + 1$ D surface (brane) embedded in a $3 + 1 +$ D space-time (bulk), Randall-Sundrum brane-word model (warped extra-dimensions) \cite{brane4, brane5} which provides a solution to the hierarchy problem, the DGP model \cite{brane6ii}, ADD model (large extra dimensions) \cite{bradd}, the thick brane model \cite{brane6iii}, and the universal extra dimension UED model \cite{brane6iiii}. The DGP model allows the cosmic accelerating expansion without introducing a cosmological constant \cite{brane7i}. However, the model is very complicated and its viability is in question \cite{brane7ii}. \par
The cosmology associated with extra dimensions has been a subject of intense study, the existence of such spatial extra dimensions can clearly alter the late-time and early-time evolution of the universe. In brane-world cosmology, a geometric alternative to the dark energy has been suggested through the extrinsic curvature of the brane which is associated to the dark energy \cite{brane8}. The phantom behavior without a big rip singularity is also an interesting feature of brane-world cosmology \cite{brane6iiii4}. The effect of one extra dimension on the properties of dark energy has been studied in \cite{brane10}. For a review on string and superstring cosmology see \cite{clever, clever2}. While most of the matter in the present universe is dark and the nature of this dark matter (DM) is unknown, many cosmological evidences provide confirmations that DM is non-baryonic. The main motivation behind the proposal of large extra dimensions \cite{bradd} was to provide a solution to the hierarchy problem between the fundamental Planck scale $M_{Pl}$ and the weak scale $M_{ew}$, the large volume extra dimension can lower $M_{Pl}$ to $M_{ew}$. \par
The UED represents a particular case of this proposal where all the standard model (SM) fields can propagate into compact extra dimensions which can help in the gauge coupling unification. This is in contrast to brane-worlds where all SM fields are confined to the $3$D brane and only gravity is free to access the extra dimension. In the UED model, the LKP is stable and could therefore still exist in the present epoch possessing all the WIMPs properties in case it is neutral and non-baryonic \cite{ued1}. Two basic assumptions of this interesting proposal are: 1- The size of the extra dimensions remains constant (stabilization of extra dimensions). 2- The cosmic evolution is governed by standard cosmology. The paper is organized as follows: In section 2, we probe the cosmological equations with $n$ universal extra dimensions through $n$-dimensional ad-hoc ansatz which satisfies recent observations. The evolution of different parameters have been studied in section 3, the final conclusion is included in section 4.

\section{Probing the universe with $n$ UED.} \label{sol}
Cosmological solutions to Einstein equations in $ 3+n+1$ D which are relevant to describe a universe with UED have been analyzed in \cite{ued1}. The setup proposed in \cite{ued1} suggests that a universe with $n$ UED can be described by a homogeneous FRW metric with two different scale factors in the $3$D and the UED:
\begin{equation} \label{metric}
ds^2=-dt^2+a^2(t)\gamma_{ij}dx^idx^j+b^2(t)\tilde{\gamma}_{pq}dy^pdy^q,
\end{equation}
where $\gamma_{ij}$ and $\bar{\gamma}_{pq}$ are maximally symmetric metrics in $3$D and $n$D, respectively. Spatial curvature is $k_a = -1, 0, 1$ in ordinary space and $k_b = -1, 0, 1$ in the UED. The energy-momentum tensor describes a homogeneous but anisotropic perfect fluid, and takes the form:
\begin{equation}
\label{emtensor}
  T^A_{\phantom{A}B} = \left(
  \begin{array}{ccc}
    -\rho & 0 & 0 \\
    0 & \gamma^i_{\phantom{i} j} p_a & 0 \\
    0 & 0 & \tilde{\gamma}^p_{\phantom{p} q} p_b
  \end{array}
  \right) \,,
\end{equation}
The equation of state (EoS) parameter is $\omega_a=\frac{p_a}{\rho}$ in ordinary space, and $\omega_b=\frac{p_b}{\rho}$ in UED where $p_a$ and $p_b$ are the pressure in ordinary space and UED respectively. The nonzero components of the Einstein equations in the background metric (\ref{metric}) can be written as
\begin{eqnarray} \label{extra}
3\left(\frac{\dot{a}}{a}\right)^2+3\frac{k_a}{a^2}+3n \frac{\dot{a}}{a}\frac{\dot{b}}{b}+\frac{n(n-1)}{2}\left[\left(\frac{\dot{b}}{b}\right)^2+\frac{k_b}{b^2}\right]=\kappa^2\rho,~~~~~~\\
2\frac{\ddot{a}}{a}+\left(\frac{\dot{a}}{a}\right)^2+\frac{k_a}{a^2}+n \frac{\ddot{b}}{b}+2n \frac{\dot{a}}{a}\frac{\dot{b}}{b}+\frac{n(n-1)}{2}\left[\left(\frac{\dot{b}}{b}\right)^2+\frac{k_b}{b^2}\right]=-\kappa^2\omega_a\rho,  \label{extra1} \\
3\frac{\ddot{a}}{a}+3\left(\frac{\dot{a}}{a}\right)^2+3\frac{k_a}{a^2}+(n-1) \frac{\ddot{b}}{b}+3(n-1) \frac{\dot{a}}{a}\frac{\dot{b}}{b}+\frac{(n-1)(n-2)}{2}\left[\left(\frac{\dot{b}}{b}\right)^2+\frac{k_b}{b^2}\right]=-\kappa^2\omega_b\rho. \label{extra2}
\end{eqnarray}
In \cite{ued1}, Einstein equations for a $ 3+n+1$ D homogeneous but anisotropic universe have been analyzed. In this study it has been found that while static extra dimensions can arise naturally in the radiation-dominated era, no solutions can allow them or allow the usual scale factor's behavior for the ordinary 3D space in the matter-dominated era. They have also investigated if there is any time evolution of the extra dimensions that corresponds  to the standard 4D cosmology, and concluded that an explicit stabilization mechanism which can reproduce standard cosmology is required. Another attempt to solve the $n$D cosmological equations (\ref{extra}-\ref{extra2}) has been introduced in \cite{ued2} by utilizing Kasner-type solution. It has been shown that, in the framework of UED, a late-time cosmological picture similar to that of the standard cosmology can be recovered in a multidimensional scenario. In the current work we investigate these $n-$dimensional cosmological equations through $n$D empirical ansatz which produces a good agreement with the standard $\Lambda$CDM in $4$D for $0<r<1$.
\begin{equation} \label{ansatz}
a_{n+4}(t)=\frac{(n+4)A}{4} \sinh^{r}(\xi t), ~~~ b_{n+4}(t)=\frac{(n+4)B}{4} \sinh^{r}(\eta t).
\end{equation}
In other words, since cosmological observations show that the present universe accelerates after an epoch
of deceleration \cite{11,rrr}, new solutions can always be explored through physically reasonable empirical scale factors which flips the sign of the deceleration parameter $q$ from positive to negative. The deceleration parameter $q$ is the most important parameter from observational viewpoint, and it carries the total effects of cosmic fluids \cite{crossing1}. Because the $4$D limit of this solution (at $n=0$) agrees with the $4$D cosmological observations, we can trust the information we get through it on cosmic evolution in UED. This $n$D empirical ansatz method is an alternative way to probe the higher dimensional cosmological equations. One basic advantage of this approach is that there are no worries on reproducing the $4$D standard cosmology. In addition, the $4$D limit of this $n$D ansatz appears in many contexts of cosmology and different gravity theories where a good agreement with observations has been obtained. It has been used in Bianchi cosmology \cite{pr} and entropy-corrected cosmology \cite{nasramri} where stable cosmological models have been obtained. The form $a(t)= \frac{a_o}{\alpha}[\sinh(t/t_o)]^{\beta}$, has been used in \cite{sen} to build a quintessence model with double exponential potential. As has been indicated in \cite{sen}, the basic motivation behind using this form was its consistency with observations where it leads to both early-time deceleration and late-time acceleration. We start by analyzing the behavior of the deceleration and jerk parameters \cite{j1,j2} for the general ansatz (\ref{ansatz}): 
\begin{equation} \label{q1}
q(t)=-\frac{\ddot{a}a}{\dot{a}^2}=\frac{-\cosh^2(\xi t)+r}{\cosh^2(\xi t)},~~~~~~j(t)=\frac{\dddot{a}}{aH^3}= 1+\frac{2r^2-3r}{\cosh^2(\xi t)}
\end{equation}
where $\dddot{a}$ is the third derivative of $a(t)$. We note that the expressions for $q$ and $j$ are independent of $n$ which means that the behavior of both parameters are the same in any number of dimensions. Since flat $\Lambda$CDM models have $j = 1$ \cite{j3}, this parameter is useful to measure the departure from the standard $\Lambda$CDM. The evolution of $j(t)$ is shown in figure \ref{F63j} for different values of $r$. The sign flipping of $q(t)$ is shown in figure \ref{F63} for different values of $r$. The cosmic transition is expected to occur at some time where $q=0$ ( or $\ddot{a}=0$). Here we get
\begin{equation}
t_{q=0}= \frac{1}{2b}\ln\left(2r-1+2\sqrt{r^2-r}\right),
\end{equation}
The current value of $q$ is expected to be around $-0.55$ \cite{sz2}. Solving (\ref{extra}), (\ref{extra1}) and (\ref{extra2}) with (\ref{ansatz}), the expressions for the pressure $p(t)$, energy density $\rho(t)$, and EoS parameter $\omega(t)$ can be written as:
\begin{eqnarray} \label{11}
\rho(t)&=& \frac{f(t)}{8\kappa^2 \sinh^2(\xi t) \sinh^2(\eta t)},\\
p_a(t)&=& \frac{g(t)}{8\kappa^2 \sinh^2(\xi t) \sinh^2(\eta t)},~~~~~~p_b(t)= \frac{h(t)}{8\kappa^2 \sinh(\xi t) \sinh^2(\eta t)}, \label{12}\\
\omega_a(t)&=&\frac{p_a}{\rho}=\frac{g(t)}{f(t)}, ~~~~~~~~~~~~~~~~~~~~~~\omega_b(t)=\frac{p_b}{\rho}=\frac{h(t)}{f(t)}. \label{13}
\end{eqnarray}
where 
\begin{eqnarray}
f(t)&=&\left(\left((n^2-n)\eta^2+6\xi^2 \right)\cosh^2(\xi t)-\eta^2 (n^2-n)\right)\cosh^2(\eta t) \\  \nonumber
 &+& 6n\xi \cosh(\xi t) \eta \cosh(\eta t) \sinh(\xi t) \sinh(\eta t) -6\xi^2\cosh^2(\xi t),
\end{eqnarray}
\begin{eqnarray}
g(t)&=&\left(\left((-n^2-n)\eta^2-6\xi^2 \right)\cosh^2(\xi t)+\eta^2 (n^2+n)+8\xi^2\right)\cosh^2(\eta t) \\  \nonumber
 &-& 4n\xi \cosh(\xi t) \eta \cosh(\eta t) \sinh(\xi t) \sinh(\eta t) + (4n\eta^2+6\xi^2) \cosh^2(\xi t)\\
&-& 4n\eta^2-8\xi^2,
\end{eqnarray}
\begin{eqnarray}
h(t)&=&\left(\left((-n^2+1)\eta^2-12\xi^2 \right)\cosh^2(\xi t)+\eta^2 (4n-4)+12\xi^2\right)\sinh(\eta t) \\  \nonumber
 &-& 6(n-1)\xi\eta \cosh(\xi t) \sinh(\eta t) \cosh(\eta t).
\end{eqnarray}
In contrast to the jerk and deceleration parameters, the quantities in (\ref{11}), (\ref{12}) and (\ref{13}) depend on the number of dimensions $n$. Therefore, we need to carefully investigate how the evolution of these quantities through cosmic time changes as the number of dimensions changes.  
\section{Dark energy evolution}
Table \ref{tap} summarizes the behavior of the energy density, pressure, and EoS parameters in normal and extra dimensions. The evolution of the energy density is always physically acceptable where $\rho \rightarrow \infty$ as $t \rightarrow 0$.
\subsection{$(+ve)\rightarrow(-ve)$ pressure transition in $3$ and $4$ D.}
The behavior of $p_a$ in $3 (n=-1)$ and $4 (n=0)$ dimensions is the same, it is positive for the early-time decelerating epoch and negative for the late-time accelerating epoch. Since positive pressure represents attractive gravity and negative pressure represents repulsive gravity, this 'positive-negative transition' in the behavior of $p_a$ gives an explanation to the cosmic 'deceleration-acceleration transition' although the reason behind the transition itself still unknown. This behavior of the \textit{positive-negative pressure transition} appears in other cosmological contexts such as cyclic universes \cite{nasrcyc} and entropy-corrected cosmology \cite{nasrent,nasramri}.  
\subsection{higher dimensions}
\begin{figure}[H] \label{tap1}
  \centering             
  \subfigure[$q$]{\label{F63}\includegraphics[width=0.29\textwidth]{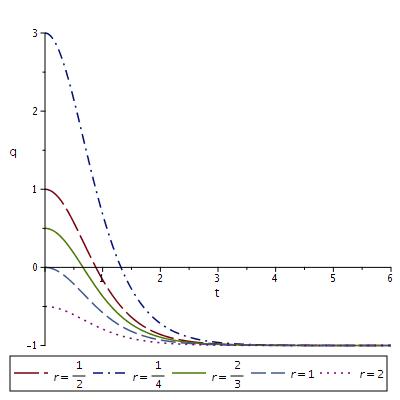}} 
	\subfigure[$j$]{\label{F63j}\includegraphics[width=0.29\textwidth]{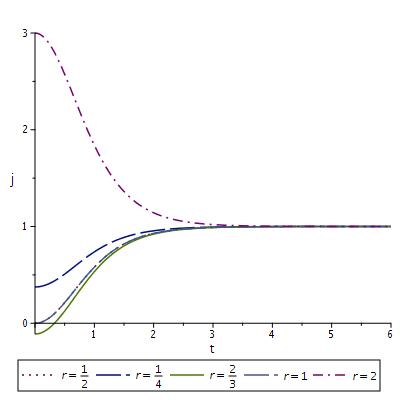}} 
	  \subfigure[$p_a$]{\label{F63a}\includegraphics[width=0.29\textwidth]{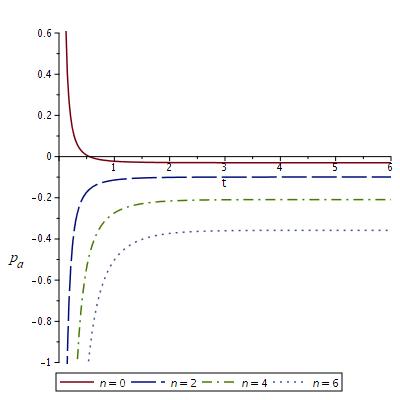}} \\
	\subfigure[$\omega_a$]{\label{F63ja}\includegraphics[width=0.29\textwidth]{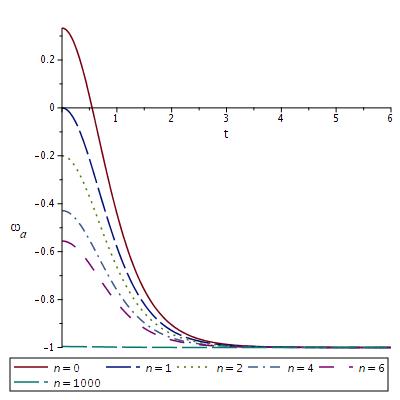}} 
	  \subfigure[$p_b$]{\label{F63s}\includegraphics[width=0.29 \textwidth]{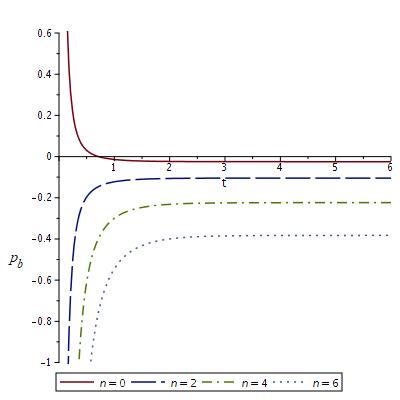}} 
	\subfigure[$\omega_b$]{\label{F63js}\includegraphics[width=0.29\textwidth]{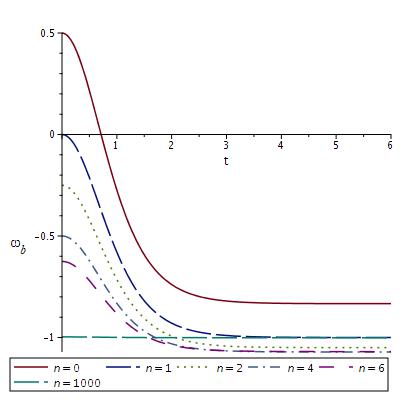}} 
  \caption{ \ref{F63} and \ref{F63j} show the deceleration and jerk parameters for different values of $r$. \ref{F63a} The behavior of $p_a$ for different $n$. \ref{F63ja} $\omega_a$ doesn't cross the $-1$ boundary, we have increased $n$ just to show that the expression of $\omega_a$ doesn't ever go behind $-1$. \ref{F63s} The pressure in UED for different $n$. \ref{F63js} $\omega_a$ crosses the $-1$ boundary in UED. }
  \label{fig:cassimir55}
\end{figure}

There's no positive-to-negative transition behavior in the pressure for higher dimensions, Table 1 shows that both $p_a$ and $p_b$ have the same behavior in $5$, $6$, $7$, $8$, $9$ and $10$ dimensions (always negative). However, the evolution of the EoS parameter in higher dimensions shows an interesting feature. While the $4$-dimensional EoS parameter $\omega_a$ doesn't cross the phantom divide line at $-1$ (no quintom behavior), the higher dimensional one  $\omega_b$ slightly crosses this line (see figure \ref{F63js}). Dark energy with EoS parameter $<-1$ is called phantom dark energy, and it violates all energy conditions \cite{phant2}. The no-go theorem in quintom cosmology (see \cite{nogo} and references therein) forbids the EoS parameter of a single perfect fluid in FRW geometry to cross the $-1$ boundary. The current model includes a single homogenous but anisotropic perfect fluid in a homogenous FRW metric with two different scale factors in the ordinary dimensions and the universal extra dimensions. While this theorem is satisfied in the ordinary $4$D (figure \ref{F63ja}), it is violated in the universal extra dimensions where the EoS parameter can slightly evolve across the $-1$ boundary (figure \ref{F63js}). In the current work, the violation of the no-go theorem in UED suggests that this theorem might not be valid in the extra dimensions of any higher-dimensional dark energy theory but, in the same time, remains valid in the ordinary four dimensions of the theory. This can strongly depend on the specific cosmological solution and the type of the extra-dimensional theory. For example, we have found that the $\omega =-1$ crossing is forbidden in both $4$D and UED for an exponential empirical ansatz of the form $a(t)=\frac{n+4}{4} e^{\lambda t^2}$ and  $b(t)=\frac{n+4}{4} e^{\alpha t^2}$. The deceleration and jerk parameters of this empirical exponential ansatz are consistent with observations with exactly the same behavior of the deceleration and jerk parameters in (\ref{q1}). The $\omega =-1$ crossing is supported by observations \cite{cro}. In the current work, we have found a cosmological solution where this crossing doesn't appear in conventional $4$D space but it happens in other dimensions. Finally, the current result is consistent with some previous works on quintom cosmology in higher dimensions. The crossing of the cosmological constant boundary by only a single scalar field in DGP brane-world has been shown in \cite{crossing1} on the contrary of the conventional $4$D cosmology. It has also been indicated in \cite{nogo} that the $5$D gravity plays a critical role in the realization of the crossing of the cosmological constant boundary while, at the same time, the deceleration parameter $q$ is in a good agreement with observations.
\begin{table}[H]\label{tap}
\centering
\tiny
    \begin{tabular}{ | p{1.9cm} | p{2.4cm} | p{2.4cm} | p{2.3cm} | p{2.3cm} | }
    \hline
         $n=$  & $0$ & $1$ & $2$ & $3$  \\ \hline
		$\rho:\rightarrow \infty$ as $t \rightarrow 0$& \checkmark & \checkmark & \checkmark &  \checkmark  \\ \hline
    $p_a:+ve\rightarrow -ve $ & \checkmark & always $-ve$ &  always $-ve$  & always $-ve$   \\ \hline
    $p_b:+ve\rightarrow -ve $ &  & always $-ve$  &  always $-ve$  & always $-ve$    \\ \hline
    $\omega_a$ & $-1 \leq \omega_a \leq \frac{1}{3}$ & $-1 \leq \omega_a \leq 0$  & $-1 \leq \omega_a \leq -0.2$ & $-1 \leq \omega_a \leq -0.33$    \\ \hline
		$\omega_b$ &  & $-1 \leq \omega_a \leq \frac{1}{3}$ & $-1.01 \leq \omega_a \leq -0.25$  & $-1.03 \leq \omega_a \leq -0.4$    \\ \hline \hline
		$n=$ & 4 & $5$ & $6$ & $-1$   \\ \hline
		$\rho:\rightarrow \infty$ as $t \rightarrow 0$ & \checkmark & \checkmark & \checkmark  & \checkmark   \\ \hline
     $p_a:+ve\rightarrow -ve $ & always $-ve$ & always $-ve$ &  always $-ve$ &  \checkmark  \\ \hline
    $p_b:+ve\rightarrow -ve $ & always $-ve$ & always $-ve$ &  always $-ve$ &  always $+ve$  \\ \hline
    $\omega_a$ & $-1 \leq \omega_a \leq -0.43$ & $-1 \leq \omega_a \leq -0.5$ & $-1 \leq \omega_a \leq -0.55$ &  $-1 \leq \omega_a \leq 1$   \\ \hline
		$\omega_b$ & $-1.07 \leq \omega_a \leq -0.5$ & $-1.07 \leq \omega_a \leq -0.55$ & $-1.07 \leq \omega_a \leq -0.62$ & $0 \leq \omega_a \leq 2$  \\ \hline \hline
    \end{tabular}
		\caption {The behavior of $\rho$, $p_a$, $p_b$, $\omega_a$ and $\omega_b$ verses cosmic time for different number of dimensions $n$.}
		\end{table}

\section{Conclusion}
The deceleration parameter $q$ is the most important parameter from observational viewpoint which carries the total effects of cosmic fluids. In the current work, we have adopted an empirical approach to solve the higher-dimensional cosmological equations which allows $q$ to be consistent with observations. The evolution of cosmic pressure and EoS parameter across the phantom divide line in ordinary $4$D $(p_a,\omega_a)$ and universal extra dimensions $(p_b,\omega_b)$ has been investigated. The deceleration and jerk parameters for the current model  are independent of the number of dimensions $n$ which means that their behaviors (which are consistent with $4$D observations) are the same regardless the number of dimensions considered. The behavior of pressure in $4$D shows a positive-to-negative transition corresponding to the deceleration-to-acceleration cosmic transition. The current model includes a single homogenous but anisotropic perfect fluid in a homogenous FRW metric with two different scale factors in the ordinary $4$D and the UED. In the conventional $4$D cosmology, there is a no-go theorem prevents the EoS parameter of a single perfect fluid in FRW geometry to cross the $-1$ boundary. In contrast to the conventional $4$D cosmology, we have found that the dark energy evolution in UED $(\omega_b)$ shows $\omega=-1$ crossing. however, the no-go theorem is still respected in $4$D where $\omega_a$ doesn't cross the cosmological constant boundary.

\end{document}